\documentclass[pra,twocolumn]{revtex4}
\usepackage{amsmath}
\usepackage{amsfonts}
\usepackage{graphicx}

\begin{document}

\title{Quantum walks in higher dimensions} 
\author{T.\ D.\ Mackay}
\author{S.\ D.\ Bartlett}
\author{L.\ T.\ Stephenson}
\author{B.\ C.\ Sanders}
\affiliation{Department of Physics, Macquarie University, Sydney, New
  South Wales 2109, Australia} 
\date{March 19, 2002}

\begin{abstract}
  We analyze the quantum walk in higher spatial dimensions and compare
  classical and quantum spreading as a function of time.  Tensor
  products of Hadamard transformations and the discrete Fourier
  transform arise as natural extensions of the ``quantum coin toss''
  in the one--dimensional walk simulation, and other illustrative
  transformations are also investigated.  We find that entanglement
  between the dimensions serves to reduce the rate of spread of the
  quantum walk.  The classical limit is obtained by introducing a
  random phase variable.
\end{abstract}
\maketitle

\section{Introduction}

Classical random walks (also known as `drunken walks') have found
practical applications in mathematics, physics and computational
science, for example in studies of diffusion, Wiener processes and
search algorithms, respectively.  Quantum physics introduces new
perspectives, such as quantum diffusion~\cite{Mil}, quantum
stochastics~\cite{Gar}, and quantum walks~\cite{adz,abnvw,aakv,moore}.
The quantum walk is particularly appealing as an intuitively
accessible model underpinning quantum diffusion and quantum
stochastics.  Remarkable properties of these quantum walks (QWs) have
been discovered; of particular interest is that the spread (standard
deviation) for the quantum walk is proportional to elapsed time $t$,
as opposed to $\sqrt{t}$ for the classical random walk; thus, the QW
offers a quadratic gain over its classical counterpart.  

Physical implementations of the quantum walk have been
proposed~\cite{adz,tm}, and possess the attractive property that they
are inherently local in the sense that the spatial state shifts by one
step along the lattice at each time step.  One potential use of the
QW is as a benchmark for assessing the non--classical performance of a
quantum computer~\cite{tm}.  It is critically important in quantum
information to develop algorithms and processes that behave in a
distinct, observably different way than any classical one; the quantum
walk is one such example.

We extend studies of QWs to a higher number of spatial dimensions and
examine the time dependence of the standard deviation, which reveals
the universal feature of a quadratic gain over the classical random
walk.  We analyze and discuss the effects of entanglement between the
different spatial degrees of freedom.  We also compare with the
equivalent classical random walk, and obtain the classical limit from
the quantum model via the introduction of a random phase variable at
each time step and performing an ensemble average.

\section{The One--Dimensional QW}

The classical random walk in one dimension describes a particle that
moves in the positive or negative direction according to the random
outcome of some unbiased binary variable (e.g., a fair coin).  The
one--dimensional lattice on which the particle moves could be infinite
or bounded (as in a circle).  We may extend this to a QW by giving the
particle an internal degree of freedom; for example, the particle may
be a spin--$1/2$ system with internal Hilbert space $\mathcal{H}_2$
and basis states $|\pm\rangle$.  The spatial state of the particle is
given by a state in a Hilbert space $\mathcal{H}_{\text{spatial}}$ of
a one--dimensional regular lattice.  Let $|i\rangle$, with $i$ an
integer, denote the state of a particle located at position $i$; the
set $\{|i\rangle\}$ forms an orthonormal basis for
$\mathcal{H}_{\text{spatial}}$.  The total state of the particle is
given by a state in the tensor product space
\begin{equation}
  \label{eq:TotalHilbertSpace}
  \mathcal{H}_T = \mathcal{H}_{\text{spatial}} \otimes 
  \mathcal{H}_2 \, .
\end{equation}
Let the particle initially be in the spatial state $|0\rangle$ (i.e.,
localized at the origin) with internal state $|+\rangle$.  To realize
the 1--D QW~\cite{abnvw}, this particle is subjected to two
alternating unitary transformations.  The first step is the Hadamard
transformation~\cite{nc},
\begin{equation} 
  \label{eq:hadamard}
  \mathbf{H} = \frac{1}{\sqrt 2 } \begin{pmatrix}
  1&1\\ 1&-1\end{pmatrix} \, ,
\end{equation}
which acts only on the internal state of the particle (i.e., on
$\mathcal{H}_2$), and transforms the initial state $|+\rangle$ into
the superposition $\tfrac{1}{\sqrt 2} (|+\rangle + |-\rangle)$.
Following this transformation, we apply a unitary operator
$\mathbf{F}$ that translates the position of the particle
\emph{conditionally} on the internal state: if the particle has
internal state $|+\rangle$, it is moved one unit to the right, and if
the internal state is $|-\rangle$, it is moved to the left, i.e.,
\begin{align}
  \label{eq:ActionOfF}
  \mathbf{F} \bigl(|i\rangle \otimes |+\rangle \bigr)
  &= |i+1\rangle \otimes |+\rangle \, , \nonumber \\
  \mathbf{F} \bigl(|i\rangle \otimes |-\rangle \bigr)
  &= |i-1\rangle \otimes |-\rangle \, . 
\end{align}
The translation does not alter the internal state, i.e., the states
$|\pm\rangle$ are internal translation eigenstates.  Since the
transformation is linear, it will transform the superposition state
$\tfrac{1}{\sqrt 2} (|+\rangle + |-\rangle)$ into a superposition
state of the particle having moved left and right.  Thus, the internal
and spatial degrees of freedom become entangled.  The Hadamard
transformation is applied again, followed by $\mathbf{F}$, and these
transformations are repeated alternately.

\begin{figure}
  \includegraphics*[width=3.25in,keepaspectratio]{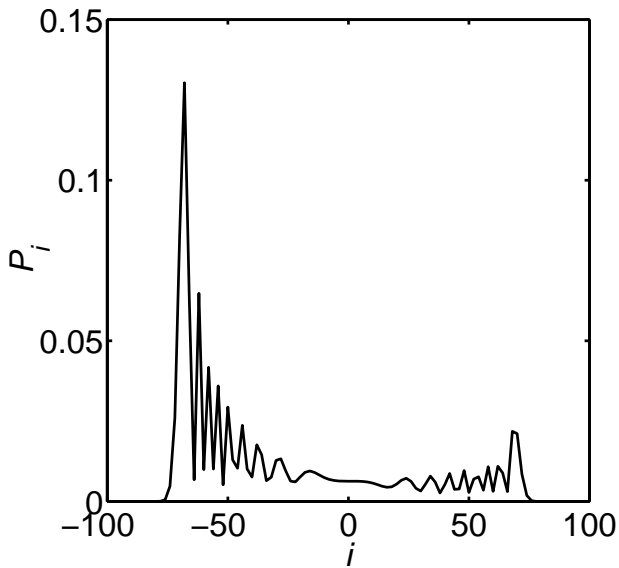}
  \caption{The probability distribution of the 1--D quantum
           walk after 100 iterations.  The internal state
           transformation used is the Hadamard transformation, and
           the initial internal state is $|-\rangle$.}
  \label{fig:h100}
\end{figure}

After $n$ iterations, the particle is in an entangled state
$|\Psi_n\rangle \in \mathcal{H}_T$.  The probability $P_i$ that the
particle will be found at the $i^{\text{th}}$ location is given by
\begin{equation}
  \label{eq:ProbabilityDistribution}
  P_i = \bigl|\bigl(\langle i| \otimes \langle +|\bigr)
  |\Psi_n\rangle\bigr|^2 + 
  \bigl|\bigl(\langle i| \otimes \langle -|\bigr)
  |\Psi_n\rangle\bigr|^2 \, .
\end{equation}
In Fig.~\ref{fig:h100}, we plot the probability distribution of this
1--D QW as a function of $i$~\cite{abnvw,aakv}.  Analytical results
are possible for the 1--D QW, and the $n\to\infty$ asymptotic
behaviour has been investigated~\cite{abnvw}.  A key feature of the
quantum walk is quantum interference, whereby two separate paths
between two nodes can interfere according to the phase difference.  In
contrast, the classical model has additive probabilities for alternate
paths.  Perhaps most interesting is the relative uniformity of the
central portion of the distribution ($-25<i<25$) and the standard
deviation of the distribution increases linearly with the number of
steps $t$; this result is in contrast to the square root dependence of
the classical random walk.  Another peculiar feature is the asymmetry
of the spatial probability distribution; this asymmetry is a
consequence of the choice of initial state.  The distribution
resulting from the initial internal state $|\psi_s\rangle =
\tfrac{1}{\sqrt 2} (|+\rangle + \text{i}|-\rangle)$ is symmetric.

\section{QW in Higher Dimensions}

The analysis of the one--dimensional walk can be extended to
higher dimensions.  We define generalizations of the Hadamard gate,
which place the internal state of the particle in superpositions of
internal translation eigenstates, plus a generalization of
$\mathbf{F}$, which moves the particle in the $d$--dimensional space
conditional on the internal state of the particle.

For a QW in $d$--dimensions, we require the particle to have an
internal state in a $2^d$--dimensional Hilbert space.  This internal
state is simply described as the state of $d$ coupled
qubits~\cite{nc}; thus, we can express the internal Hilbert space
$\mathcal{H}_{\text{int}}$ as
\begin{equation}
  \label{eq:GeneralInternalHilbertSpace}
  \mathcal{H}_{\text{int}} = \mathcal{H}_2 \otimes \mathcal{H}_2 \otimes
  \cdots \otimes \mathcal{H}_2 = \otimes^d \mathcal{H}_2 \, ,
\end{equation}
and give a basis for internal states in binary notation as
\begin{equation}
  \label{eq:GeneralInternalBasisStates}
  |\epsilon_1\epsilon_2\ldots\epsilon_d\rangle = 
  |\epsilon_1\rangle \otimes |\epsilon_2\rangle
  \otimes \ldots \otimes|\epsilon_d\rangle \, ,
\end{equation}
where $\epsilon_i = \pm$.  The state of the $i^{\text{th}}$ qubit
(with basis $|\pm\rangle$) will determine the direction (positive or
negative) that the particle moves in the $i^{\text{th}}$ dimension.
That is, we define a translation operator $\mathbf{F}$ which
translates the state of the particle by one unit in every dimension:
the direction in the $i^{\text{th}}$ dimension is conditional on the
state of the $i^{\text{th}}$ qubit.  The internal translation
eigenstates are those given in
Eq.~(\ref{eq:GeneralInternalBasisStates}).

For the 1--D QW, the quantum analogue of the classical ``coin--flip''
was the application of the Hadamard transformation of
Eq.~(\ref{eq:hadamard}).  This transformation maps an internal
translation eigenstate of the translation operator $\mathbf{F}$
(either $|+\rangle$ or $|-\rangle$) into an equally weighted
superposition of the two.  The choice of phases in this transformation
was to some extent arbitrary; the Hadamard transformation represents a
choice with real entries.

For the $d$--dimensional QW, there exists a wide variety of unitary
transformations on the internal state that could be used as a
generalization of the Hadamard transformation for the 1--D case.  One
obvious generalization would be to apply a Hadamard transformation
$\mathbf{H}$ to each qubit in the decomposition of
Eq.~(\ref{eq:GeneralInternalHilbertSpace}); i.e., the transformation
\begin{equation}
  \label{eq:TensorHadamard}
  \mathbf{H}_d = \mathbf{H} \otimes \mathbf{H} \otimes \ldots \otimes
  \mathbf{H} \, .
\end{equation}
This internal transformation is \emph{separable}, in the sense that it
does not produce entanglement between the spatial degrees of freedom.
This choice could be viewed as the quantum analogue of using $d$
independent coin tosses, one for each spatial dimension.

Another obvious generalization, which is not separable and does
produce entanglement between spatial degrees of freedom, is the
$2^d$--dimensional discrete Fourier transform (DFT) $\mathbf{D}_d$,
defined as follows.  Expressing the basis of
Eq.~(\ref{eq:GeneralInternalBasisStates}) as labelled by its numerical
value $\{ |\mu\rangle,\, \mu=0,1,\ldots, 2^d-1 \}$, the DFT acts on
this basis as
\begin{equation}
  \label{eq:DFT}
  \mathbf{D}_d |\mu\rangle = \frac{1}{\sqrt{2^d}}\sum_{\nu=0}^{2^d-1}
  e^{2\pi\text{i}\mu\nu/{2^d}}|\nu\rangle \, .
\end{equation}
Note that the Hadamard transformation is the $d=1$ discrete Fourier
transform $\mathbf{D}_1$.  As the Hadamard transformation does for the
1--D case, this DFT transforms any internal translation eigenstate
into an equally weighted superposition of all the eigenstates.  Unlike
the tensor product of Hadamard transformations, it is non--separable
and highly entangles the different internal qubits.  Although this
internal transformation can also be viewed as a quantum analogue of
$d$ independent coin tosses, this entanglement between the spatial
degrees of freedom is a genuinely quantum effect.

The DFT transformation represents a natural choice for the phase
relationship between the translation eigenstates of the
superpositions.  However, this choice of phases is arbitrary, and we
may consider other choices, which will have a different effect on the
QW.  We also investigate another internal state transformation (the
Grover operator~\cite{nc}) that also produces an equally weighted
superposition is the transformation (defined on the same basis as used
above)
\begin{equation} 
  \label{eq:grover}
  \mathbf{G}_d|\mu\rangle =\frac{1}{\sqrt{2^d}} \Bigl( -2|\mu\rangle 
  + \sum_{\nu=0}^{2^d-1} |\nu\rangle \Bigr) \, .
\end{equation}
This choice, like the Hadamard transformation, possesses only real
entries.

There are, of course, an infinite variety of other non--separable
choices for the internal transformation by employing different phase
relationships.  Also, a bias could be introduced into the
transformation, which would give an \emph{un}equally weighted
superposition of translation eigenstates; however, we consider only
unbiased transformations here.

One of the remarkable properties of the 1--D QW is that, unlike its
classical counterpart, it can produce an asymmetric distribution.
Note, however, that with appropriate initial conditions (such as the
state $|\psi_s\rangle = \tfrac{1}{\sqrt 2} (|+\rangle +
\text{i}|-\rangle)$) a symmetric distribution is obtained.  It is of
interest to question what effect the initial conditions will have on
the higher--dimensional QWs.  (Note that a symmetric distribution can
always be obtained by averaging over initial conditions.)

\section{Calculations of QWs}

\begin{figure}
  \includegraphics*[width=2.75in,height=2.75in]{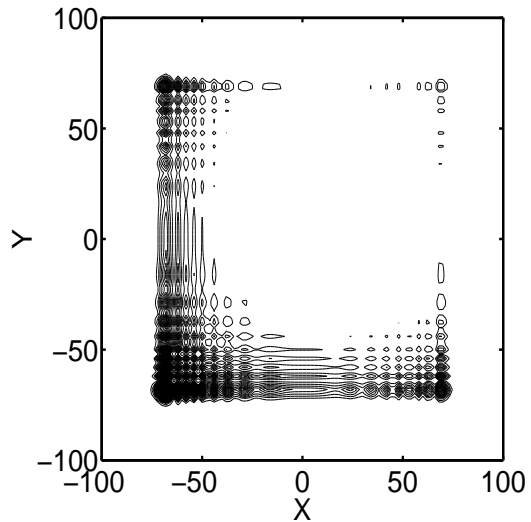}
  \caption{Probability distribution of the 2--D quantum
    walk using the separable internal transformation ${\bf
      H}\otimes{\bf H}$ over 100 iterations, with initial condition
    given by $|-\rangle\otimes|-\rangle$.}
  \label{fig:hxh100}
\end{figure}

We begin our analysis with the straightforward generalization to
higher dimensions of using the Hadamard transformation on each qubit.
Fig.~\ref{fig:h100} and Fig.~\ref{fig:hxh100} show simulation results
for the Hadamard walk both
in one--dimension and a tensor product $\mathbf{H}\otimes\mathbf{H}$
for two--dimensions respectively.  The initial condition for the
internal state was chosen to be the separable state composed of all
qubits in the $|-\rangle$ state, which leads to an asymmetric
probability distribution.

For the case of separable transformations with separable initial
conditions, the different spatial dimensions behave independently;
thus, the variance can be expressed in terms of the one--dimensional
case.
\begin{figure}
  \includegraphics*[width=3.25in,keepaspectratio]{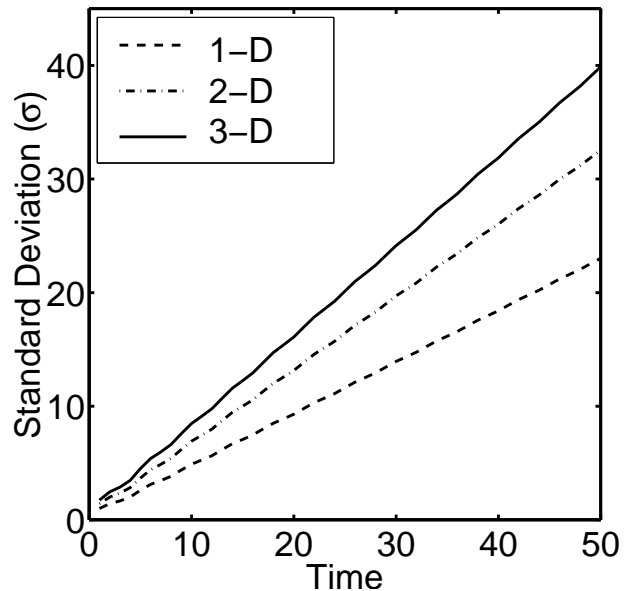}
  \caption{Time dependence of standard deviation for the series ${\bf
      H}$, ${\bf H}\otimes{\bf H}$ and ${\bf H}\otimes{\bf
      H}\otimes{\bf H}$ with initial state given by $\otimes^d
    |-\rangle$.}
  \label{fig:vartimehxh}
\end{figure}
For example, consider the family
\begin{equation}
  \label{eq:FamilyOfHadamards}
  {\bf H},\, {\bf H}\otimes{\bf H},\,
  {\bf H}\otimes{\bf H}\otimes{\bf H}, \ldots ; 
\end{equation}
the time dependence of the standard deviation for these walks is
plotted in Fig.~\ref{fig:vartimehxh}, and the corresponding slopes
$\Delta\sigma/\Delta t$ are presented in Table~\ref{tab:vartimehxh}.
We observe that
\begin{equation}
  \label{eq:DependenceOfSigma}
  (\frac{\Delta\sigma_1}{\Delta t}, \frac{\Delta\sigma_2}{\Delta t},
  \frac{\Delta\sigma_3}{\Delta t}, \ldots) = 
  (\frac{\Delta\sigma_1}{\Delta t}, \sqrt{2}
  \frac{\Delta\sigma_1}{\Delta t}, \sqrt{3}
  \frac{\Delta\sigma_1}{\Delta t}, \ldots) \, , 
\end{equation}
where $\sigma_d$ is the standard deviation for the $d$--dimensional
QW, as expected for independent distributions. Also, by calculating a
$d$--dimensional QW using this separable internal transformation and
projecting the state onto the Hilbert space for any one dimension, the
state of the 1--D QW is recovered.  Again, this property illustrates
that the different spatial dimensions are independent.
\begin{table}[ht]
  \begin{tabular}{lll}
    {\bf Transformation}&
    {$\Delta\sigma/\Delta t$}&$(\sqrt{d})\Delta\sigma_1/\Delta t$\\
    \hline
    ${\bf H}$&$0.4544\pm0.0012$&$0.4544\pm0.0012$\\
    ${\bf H} \otimes {\bf
    H}$&$0.6427\pm0.0017$&$0.6427\pm0.0017$\\ 
    ${\bf H}\otimes{\bf H}\otimes{\bf
    H}$&$0.7871\pm0.0021$&$0.7871\pm0.0021$\\ 
  \end{tabular}
  \caption{The slope of the standard deviation as a function of time for
    the family $({\bf H},\,{\bf H}\otimes{\bf H},\,
    {\bf H}\otimes{\bf H}\otimes{\bf H}, \ldots)$.  The slope 
    $\Delta\sigma/\Delta t$ is found by linear regression
    of data points where $t\geq10$ (such as
    to allow stabilization of irregularities caused by initial
    condition).  $\sigma_1$ refers to the 1--D case.}
  \label{tab:vartimehxh}
\end{table}
Analytical results for these QWs follow from the 1--D case in a
straightforward manner. 

\begin{figure}
  \includegraphics*[width=2.75in,height=2.75in]{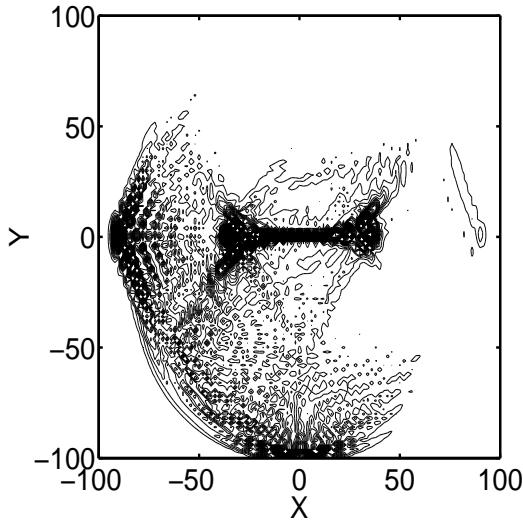}
        \caption{Probability distribution for the quantum walk
          using the $d=2$ DFT (${\bf D}_2$) over 100
          iterations, with initial condition given by
          $|-\rangle\otimes|-\rangle$.}
  \label{fig:dft2d} 
\end{figure}

We now consider the behaviour of higher--dimensional QWs that possess
entanglement between the spatial degrees of freedom, i.e., QWs that
have non--separable internal transformations, such as the DFT of
Eq.~(\ref{eq:DFT}).  Fig.~\ref{fig:dft2d} shows the spatial
probability distribution
of the QW with internal transformation given by the $d=2$ discrete
Fourier transform $\mathbf{D}_2$; note that this distribution is
distinct from that of the $\mathbf{H}\otimes\mathbf{H}$ QW.  In
particular, it has the feature that the density of the distribution is
significant near the origin, in constrast to the separable
$\mathbf{H}\otimes\mathbf{H}$ QW which possesses only average density
at the origin.  Note also that it is asymmetric for the initial
condition $|-\rangle\otimes|-\rangle$; the asymmetry appears to be a
general property of the higher--dimensional QWs as it is for the 1--D
case.

\begin{figure}
  \includegraphics*[width=3.25in,keepaspectratio]{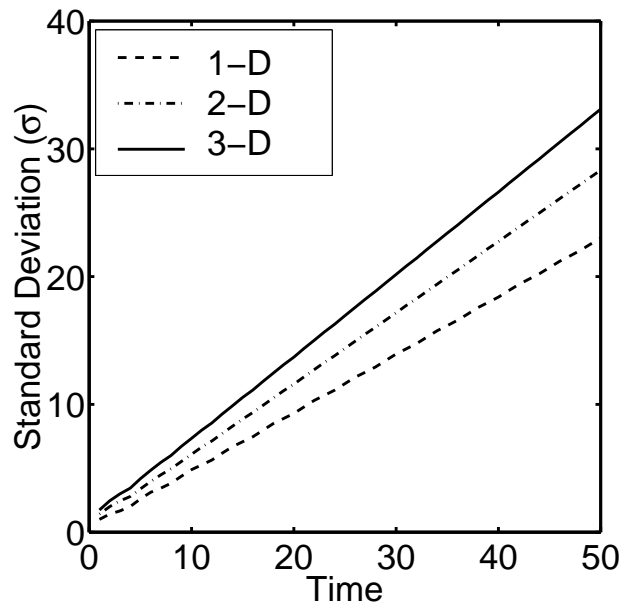}
  \caption{Time dependence of the standard deviation for the
    ${\bf D}_d$ DFT series with initial state given by $\otimes^d
    |-\rangle$.  Details are given in Table~\ref{tab:vartimedft}.}
  \label{fig:vartimedft}
\end{figure}

The time dependence of the standard deviations for the $d=1,2,3$ DFT
walks are plotted in Fig.~\ref{fig:vartimedft} and
summarized in Table~\ref{tab:vartimedft}.  In contrast to the
separable case, the trend observed in the DFT family is
$\Delta\sigma_d/\Delta t = \sqrt{(d+1)/2}\Delta\sigma_1/\Delta t$.
This trend is in agreement with the three calculations ($d=1,2,3$).
For the family of DFT walks, the standard deviation grows linearly
with time, but the slope is less than that for the separable case (the
tensor products of Hadamard transformations); see
Table~\ref{tab:vartimedft}.  This suggests that the entanglement
between the spatial degrees of freedom serves to reduce the rate of
spread.
\begin{table}[ht]
  \begin{tabular}{lll}
    \textbf{Transformation}&
    {$\Delta\sigma/\Delta t$}&{$\sqrt{(d+1)/2}\Delta\sigma_1/\Delta t$}\\
    \hline
    ${\bf D}_1$ (${\bf H}$)&$0.4544\pm0.0012$&$0.4544\pm0.0012$\\
    ${\bf D}_2$ &$0.5569\pm0.0006$&$0.5565\pm0.0015$\\
    ${\bf D}_3$ &$0.6449\pm0.0007$&$0.6426\pm0.0017$\\
  \end{tabular}
  \caption{Slope of the standard deviation as a function of time, and
    comparison to the suggested pattern.  $\Delta\sigma/\Delta t$ is
    the slope found by linear regression of data points where
    $t\geq10$ (such as to allow stabilization of irregularities caused
    by initial condition).} 
  \label{tab:vartimedft}
\end{table}

\begin{figure}
  \includegraphics*[width=2.75in,height=2.75in]{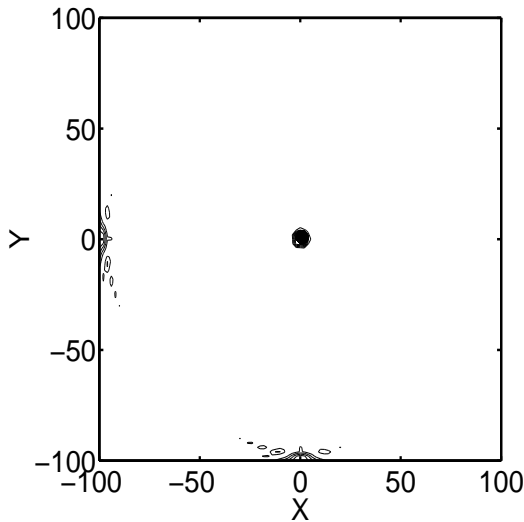}
        \caption{Probability distribution of the 2--D quantum
            walk with internal 
          transformation given by $\mathbf{G}$ (see
          Eq.~\ref{eq:grover}) over 100 iterations with initial
          condition $|-\rangle\otimes|-\rangle$.}
  \label{fig:grover}
\end{figure}

Choosing different relative phases in the internal state
transformation can lead to vastly different distributions.
Fig.~\ref{fig:grover} shows the results of using the internal
transformation $\mathbf{G}$ of Eq.~(\ref{eq:grover}).  This
distribution is charaterized by its marked localization at the centre,
as well as possessing peaks at the ``maximum distance'' attainable in
the number of iterations (100 units from the origin).

Note that the time dependence of variance for the non--separable 2--D
transformations ($\mathbf{D}_2$, $\mathbf{G}_2$) are quite similar,
although the probability density functions are quite different in
appearance.  (See Table~\ref{tab:vartime2d}.)  The choice of initial
condition does not appear to have a significant effect on the time
dependence of the standard deviation.
\begin{table}[ht]
  \begin{tabular}{lll}
    \textbf{Transformation} & \textbf{Initial State} &
    {$\Delta\sigma/\Delta t$}\\ 
    \hline
    $({\bf H} \otimes {\bf
      H})$&$|-\rangle\otimes|-\rangle$&$0.6427\pm0.0017$\\ 
    ${\bf D}_2$ &$|-\rangle\otimes|-\rangle$&$0.5569\pm0.0006$\\
    ${\bf D}_2$ &$|+\rangle\otimes|+\rangle$&$0.5569\pm0.0006$\\
    ${\bf D}_2$ &$|\psi_s\rangle \otimes |\psi_s\rangle$&$0.6234\pm0.0005$\\
    ${\bf D}_2$ &$|\psi_-\rangle$&$0.6009\pm0.0006$\\
    $\mathbf{G}$ (see
    Eq.~\ref{eq:grover})&$|-\rangle\otimes|-\rangle$&$0.5418\pm0.0020$\\ 
    $\mathbf{G}$&$|+\rangle\otimes|+\rangle$&$0.5418\pm0.0020$\\
    $\mathbf{G}$&$|\psi_s\rangle \otimes |\psi_s\rangle$&$0.5988\pm0.0006$\\
    $\mathbf{G}$&$|\psi_-\rangle$&$0.5440\pm0.0008$\\
  \end{tabular}
  \caption{Slope of the standard deviation as a function of time for
      various 2--D transformations and initial conditions.
      $\Delta\sigma/\Delta t$ is the slope
      found by linear regression of data
      points where $t\geq10$ (to allow stabilization of
      irregularities caused by initial
      condition). Here $|\psi_s\rangle=\tfrac{1}{\sqrt{2}}(|+\rangle+
      \rm{i}|-\rangle)$ is the state that produces the 1--D symmetric
        distribution, and $|\psi_-\rangle = \tfrac{1}{\sqrt{2}}(|+\rangle
        \otimes|-\rangle - |-\rangle\otimes|+\rangle)$ is the entangled
        singlet state.}
  \label{tab:vartime2d}
\end{table}

\section{Obtaining the Classical Random Walk from the Quantum Model}

A classical distribution can be obtained from the quantum model by
introducing a random element into the transformation at each time
step.  As shown previously, the ``quantum'' behaviour of the QW is
due to the phase relationship (interference) between the separate
paths of the walk.  By adding a random element to the phase and
averaging over many trials, we show that the quantum inteference can
be made to disappear and that the distribution of the classical random
walk is regained.  The introduction of this random phase is an example
of decoherence.

Let us first investigate the one--dimensional case.  In the internal
translation eigenstate basis $|\pm\rangle$, the unitary operator that
transforms the relative phase between these states is
\begin{equation} 
  \label{eq:ranphase}
  \mathbf{R}(\beta) = e^{\frac{\text{i}}{2} \beta \hat{\sigma}_z} 
  = \begin{pmatrix} {e^{\text{i}\beta/2}} & 0 \\
  0 & {e^{-\text{i}\beta/2}} \end{pmatrix}\, ,
\end{equation}
where $\hat{\sigma}_z = \left(\begin{smallmatrix} 1 & 0 \\ 0 & -1
  \end{smallmatrix} \right)$ is the Pauli spin matrix, and $\beta \in
[0,2\pi)$.  We then consider a QW where the phase between the
$|+\rangle$ and $|-\rangle$ states is randomly selected at each
interval from a uniform prior distribution over $[0,2\pi)$.  Rather
than applying the Hadamard transformation as the internal
transformation, we apply
\begin{equation} 
  \label{eq:ranphaseHadamard}
  \mathbf{H}(\beta) = \mathbf{R}(\beta)\, \mathbf{H}\, 
  \mathbf{R}(\beta)^{-1} 
  = \begin{pmatrix} 1 & {e^{\text{i}\beta}} \\
  {e^{-\text{i}\beta}} & -1 \end{pmatrix}\, ,
\end{equation}
with a phase $\beta$ chosen randomly from the set $[0,2\pi)$ at each
time step.  The resulting distribution has a stardard deviation
comparable to that of the corresponding binomial distribution, but
exhibits strong interference effects.  By averaging oven many trials,
the distribution rapidly converges to the binomial distribution.

These results for the one--dimensional case can easily be generalized
to higher dimensions.  For the separable $d$--dimensional QW, the
generalization is straightforward: one simply replaces each Hadamard
transformation in the tensor product with a random $\mathbf{H}(\beta)$
at each step.  The separability ensures that the resulting walk is
equivalent to the 1--D walk in each dimension.

For non--separable internal transformations such as the DFT, a
straightforward extension is to apply
\begin{equation}
  \label{eq:ranphasehigherd}
  (\mathbf{R}_1(\beta_1) \otimes \cdots \otimes \mathbf{R}_d(\beta_d))\,
  \mathbf{D}_d\, (\mathbf{R}_d(\beta_d)^{-1} \otimes \cdots \otimes
  \mathbf{R}_1(\beta_1)^{-1}) \, ,
\end{equation}
where $\beta_1,\ldots,\beta_d$ are random phases, each from the set
$[0,2\pi)$.  That is, an independent random phase is added for each
dimension (qubit).  Again, by averaging 400 walks of 50 iterations
each, we obtained the 2--D binomial distribution to a high degree of
confidence.

\section{Conclusions and Discussion}

We present here a framework for calculating and analyzing quantum
walks in higher dimensions.  The generalization of these walks
beyond one dimension gives a wide variety of choice for the phases
involved in the ``quantum coin toss''.  We discuss the role of
entanglement between the different spatial degrees of freedom as a
possible non--classical property of the higher dimensional QWs.
As different choices lead to different spatial probability
distributions, it may be that specific unitary transformations of the
internal Hilbert space are particularly well suited for certain
computational tasks.

As with the one--dimensional QW, the increased rate of spread (given
by the linear dependence of the standard deviation on time) is present
in the higher dimensional walks.  This property may be particularly
valuable for classical random walk based algorithms, such as quantum
searches.  We show that entanglement between the spatial degrees of
freedom reduces the slope of this linear growth but not the linear
dependence on $t$.  These results are shown to be independent of the
initial internal state in the cases investigated.

We show that the classical distribution can be obtained from the QW
by introducing an internal transformation with a random phase and then
averaging over many trials.  This result is expected; the quantum
behaviour of the QW is due to interference effects between the phases
of different paths.  For higher dimensional QW, more random
parameters (one for each spatial dimension) are needed.

\begin{acknowledgments}
  This project has been supported by an Australian Research Council
  Large Grant and a Macquarie University Research Grant.  SDB
  acknowledges the support of a Macquarie University Research
  Fellowship.  We acknowledge helpful discussions with G.\ J.\ 
  Milburn, T.\ E.\ Freeman, T.\ Rudolph and B.\ C.\ Travaglione.
\end{acknowledgments}

\end{document}